\documentclass[a4paper,11pt]{article}
\usepackage{pos}

\title{Charmed hadron production in high-energy nuclear collisions}

\author*[a]{A. Beraudo}
\author[a]{A. De Pace}
\author[a]{M. Monteno}
\author[a]{M. Nardi}
\author[a]{D. Pablos}
\author[a]{F. Prino}

\affiliation[a]{INFN - Sezione di Torino,\\
  Via Pietro Giuria 1, 10125 Torino, Italy}

\emailAdd{beraudo@to.infn.it}

\abstract{We present a new model for the description of heavy-flavor hadronization in high-energy nuclear (and possibly hadronic) collisions, where the process takes place not in the vacuum, but in the presence of other color charges.
We explore its effect on the charmed hadron yields and kinematic distributions once the latter is applied at the end of transport calculations used to simulate the propagation of heavy quarks in the deconfined fireball produced in nuclear collisions. The model is based on the formation of color-singlet clusters through the recombination of charm quarks with light antiquarks or diquarks from the same fluid cell. This local mechanism of color neutralization leads to a strong space-momentum correlation, which provides a substantial enhancement of charmed baryon production -- with respect to expectations based on $e^+e^-$ collisions --  and of the collective flow of all charmed hadrons. We also discuss the similarities between our model and recently developed mechanisms implemented in QCD event generators to simulate medium corrections to hadronization in the presence of other nearby color charges.}

\FullConference{%
  41st International Conference on High Energy physics - ICHEP2022\\
  6-13 July, 2022\\
  Bologna, Italy
}

\begin{document}
\maketitle

\section{Introduction}
The diffusion of Brownian particles can be used to access microscopic properties of the medium in which their propagation takes place. Historically, the diffusion of small grains in water was used by Perrin to prove the atomic structure of matter and to provide a first estimate of the Avogadro number, finding ${\cal N}_A\approx 5.5-7.2\cdot 10^{23}$~\cite{Perrin:1926}. Nowadays, one of the goals of heavy-ion collisions is to exploit the \emph{relativistic} Brownian motion of heavy quarks to get an estimate of comparable accuracy of medium properties like the \emph{momentum broadening coefficient} $\kappa$. An important difference in this case is that the nature of the Brownian particle changes during its propagation, since after a few fm/c it undergoes hadronization. This introduces a source of systematic uncertainty in the extraction of transport coefficients; on the other hand it can be considered an issue of interest in itself, in particular to study how hadronization changes in the presence of a medium acting as a color reservoir. This, in the case of charm quarks, will be the subject of our study, carried out to provide an interpretation of non-trivial results concerning heavy-flavor hadrochemistry in heavy-ion collisions, but with the potential to shed light also on analogous puzzling findings obtained in smaller systems~\cite{ALICE:2020wfu}.
\section{The model}\label{Sec:model}
\begin{figure*}
\centering
\includegraphics[clip,width=0.48\textwidth]{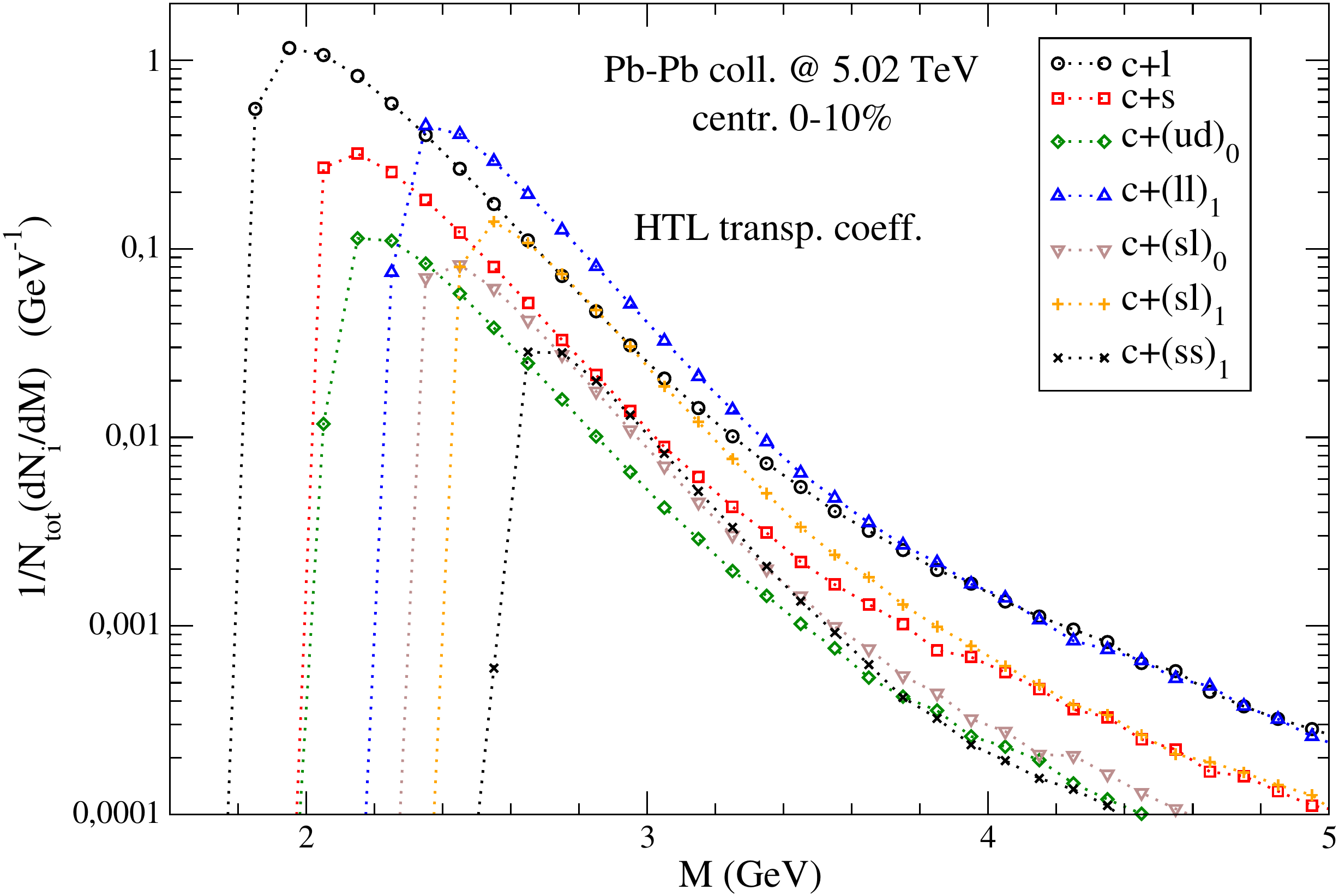}
\includegraphics[clip,width=0.48\textwidth]{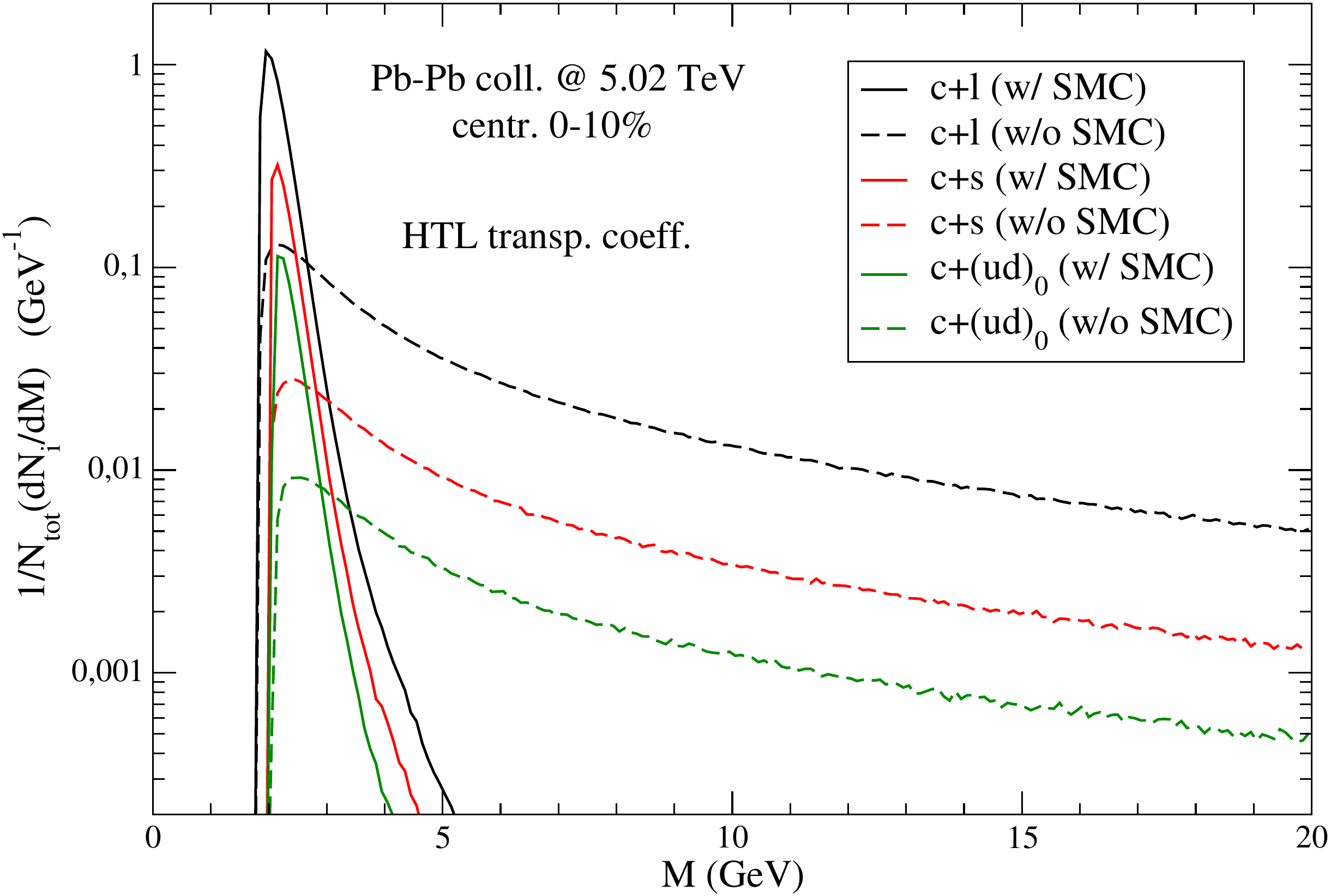}
\caption{Left panel: the invariant-mass distribution of the $Q\overline q$ and $Q(qq)$ clusters on the hadronization hypersurface in the different isospin, strangeness and spin channels. Right panel: cluster invariant-mass distributions with recombination occurring locally (continuous curves, default implementation), with strong correlation between the parton momenta and position, or non-locally (dashed curves), with no space-momentum correlation. All results refer to central Pb-Pb collisions at $\sqrt{s_{\rm NN}}\!=\!5.02$ TeV, with charm quark propagation before hadronization simulated using weak-coupling (HTL) transport coefficients.}
\label{fig:Mdistr}
\end{figure*}
Any hadronization model must start clustering colored partons into color-singlet structures which will give rise to the final hadrons. In proton-proton collisions partons are taken from the hard process, from the shower stage, from the underlying event and from the beam remnants. In heavy-ion collisions recombining partons are taken from the hot deconfined plasma produced in the collision and they are \emph{close in space} (the Debye radius sets the maximum distance at which two colored partons are correlated): since the fireball undergoes a collective expansion, this last detail has deep phenomenological consequences. 

We now briefly describe our model for in-medium charm hadronization. For more details we refer the reader to our original publication~\cite{Beraudo:2022dpz}.
Once a $c$ quark, during its stochastic propagation through the fireball, reaches a fluid cell at $T_H=155$ MeV it recombines with a
light antiquark or diquark from the same fluid element. Both the species and the momentum of the medium particle are sampled assuming they obey a thermal distribution in the local rest frame (LRF) of the fluid. The thermal particle is then boosted to the laboratory frame and recombined with the charm quark, leading to the formation of the cluster ${\cal C}$. Since the $c$ quark and the thermal particle are taken from the same fluid cell undergoing a collective flow, typically their momenta are quite collinear and correlated to the position they occupy (Space-Momentum Correlation). This leads to the production of quite low invariant-mass clusters, as one can see from the left panel of Fig.~(\ref{fig:Mdistr}). We also check that, suppressing this SMC by randomly redistributing the HQ's on the decoupling hypersurface (see right panel of Fig.~\ref{fig:Mdistr}), the production of heavier clusters is favored.
As in Herwig~\cite{Webber:1983if}, one treats differently light and heavy clusters. Clusters with $M_{\cal C}<M_{\rm max}\approx 4$ GeV undergo an isotropic two-body decay in their LRF, giving rise to a charmed hadron and a second soft particle; the very rare heavier clusters are fragmented into multiple hadrons as Lund strings~\cite{Andersson:1983ia}.
\section{Results}\label{Sec:results}
\begin{figure*}
\centering
        \includegraphics[clip,width=0.98\textwidth]{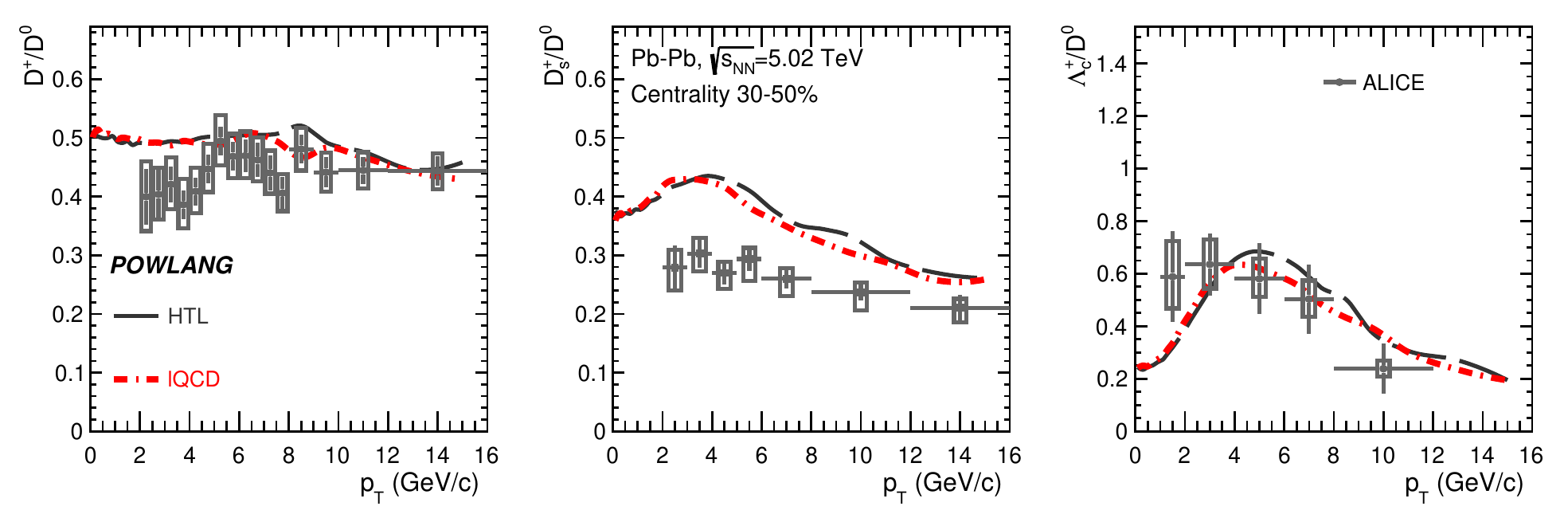}
\caption{Predictions for the relative yields of charmed hadrons (relative to $D^0$ mesons) in semi-central Pb-Pb collisions at $\sqrt{s_{\rm NN}}\!=\!5.02$ TeV for different transport coefficients compared to recent ALICE data~\cite{ALICE:2021bib,ALICE:2021kfc,ALICE:2021rxa}.}\label{Fig:D-Ds-Lc}
\end{figure*}
Interfacing our new hadronization model to numerical simulations of heavy-quark transport in the deconfined fireball one can obtain a satisfactory description of important and partially unexpected experimental findings, in particular the strong enhancement of charmed-baryon production recently observed in heavy-ion~\cite{ALICE:2021bib} and also in proton-proton collisions~\cite{ALICE:2020wfu}, which cannot be explained by any hadronization model tuned to reproduce $e^+e^-$ data.  
An example of our results is given in Fig.~\ref{Fig:D-Ds-Lc}, where we plot as a function of $p_T$ the $D^+/D^0$, $D_s^+/D^0$ and $\Lambda_c^+/D^0$ ratios in Pb-Pb collisions at $\sqrt{s_{\rm NN}}=5.02$ TeV, comparing our predictions to recent ALICE data~\cite{ALICE:2021bib,ALICE:2021kfc,ALICE:2021rxa}. Notice that if hadronization occurred as in $e^+e^-$ collisions one would get $\Lambda_c^+/D^0\approx 0.1$. As one can see the strong enhancement of the $\Lambda_c^+/D^0$ ratio for intermediate values of $p_T$ is correctly reproduced. Analogous results are obtained when our model is compared to STAR data for Au-Au collisions at $\sqrt{s_{\rm NN}}=200$ GeV~\cite{STAR:2019ank}.

\begin{figure*}
\centering
        \includegraphics[clip,height=6cm]{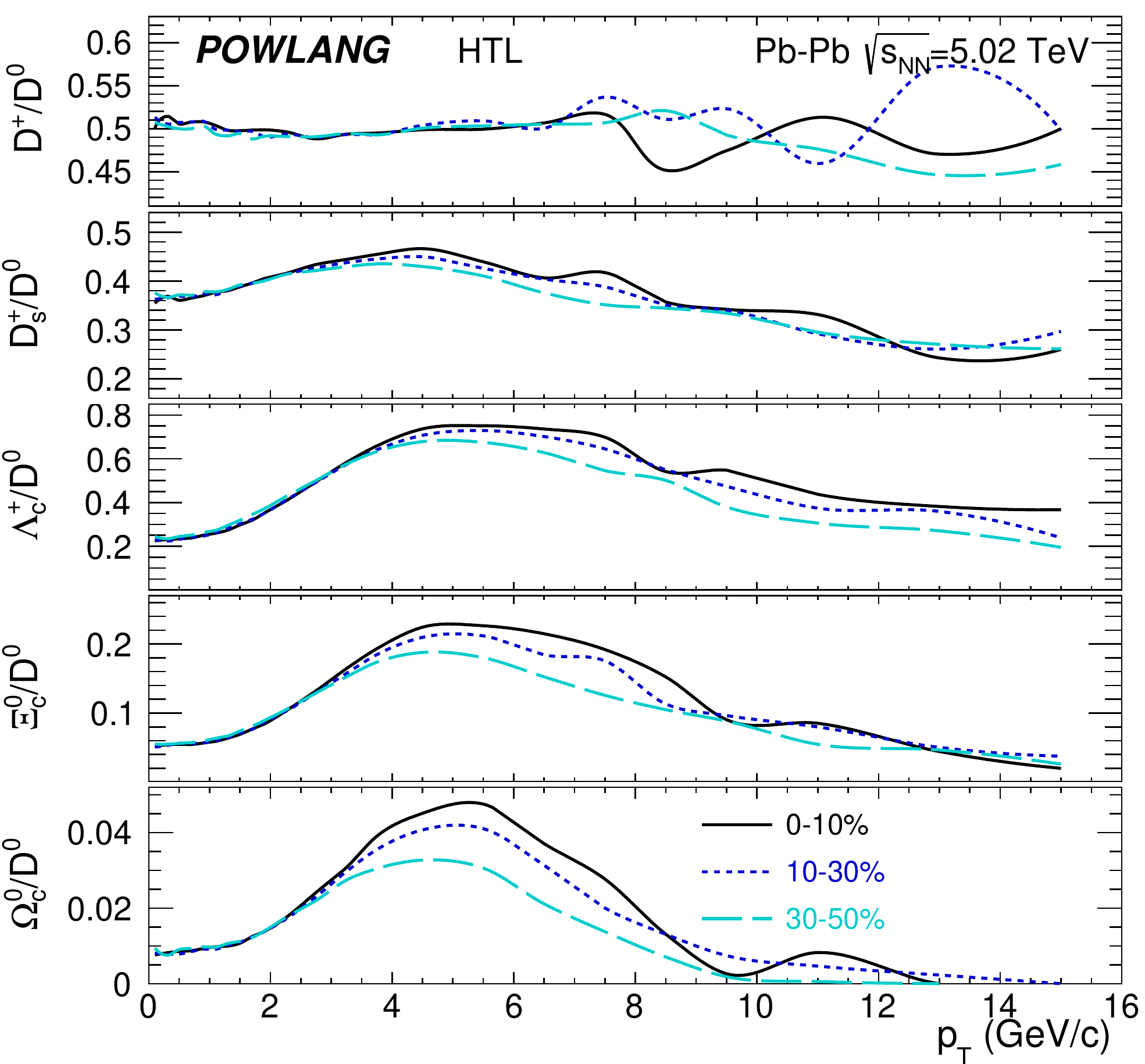}
        \includegraphics[clip,height=6cm]{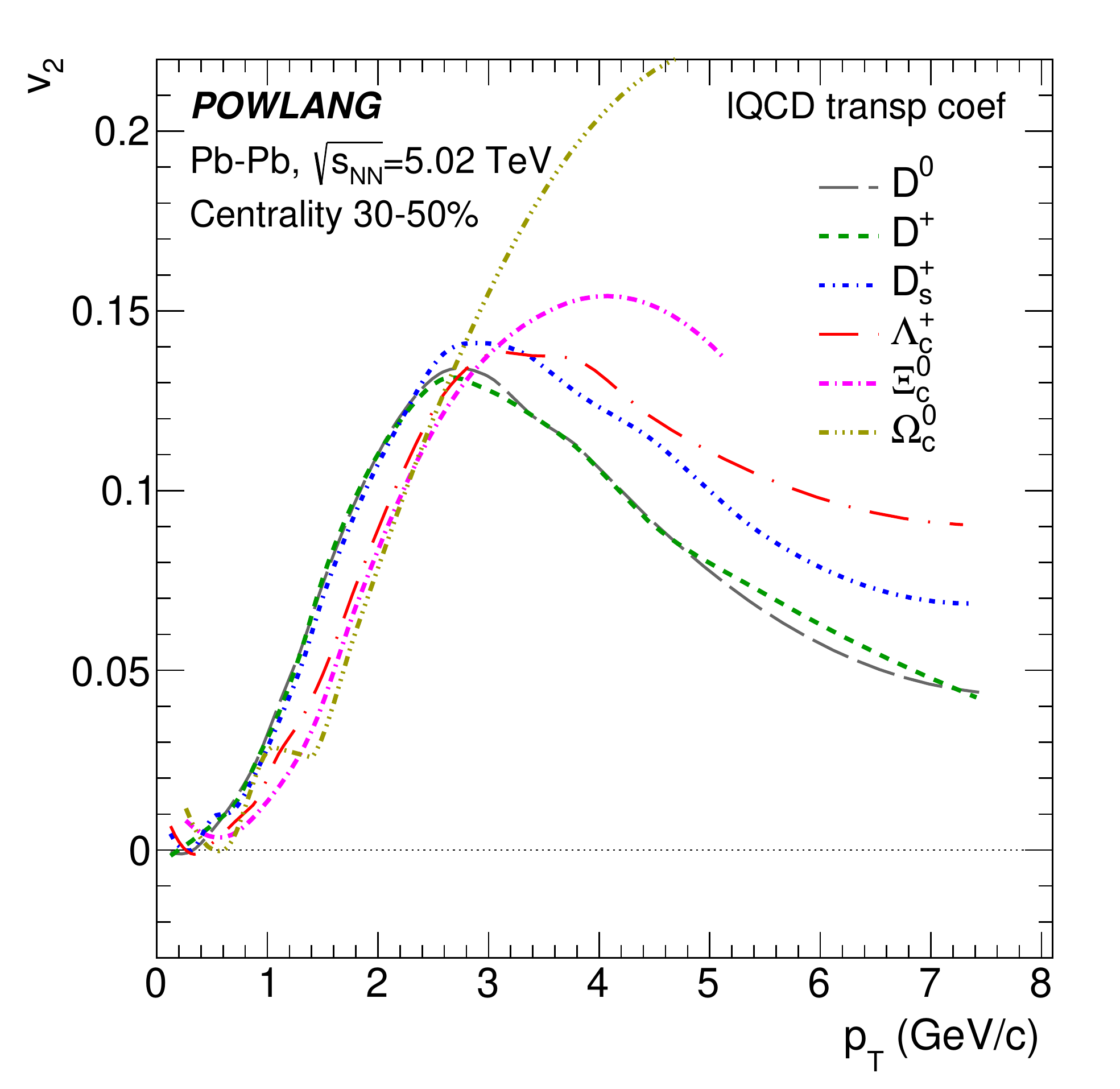}
\caption{Charmed hadron ratios (left panel, with weak-coupling transport coefficients) and elliptic flow (right panel, with l-QCD transport coefficients) as a function of $p_T$ in Pb-Pb collisions at $\sqrt{s_{\rm NN}}\!=\!5.02$ TeV.}\label{Fig:hc-vs-pt}
\end{figure*}
In Fig.~\ref{Fig:hc-vs-pt} we extend our analysis to the yields and kinematic distributions of all ground-state charmed hadrons obtained with our new hadronization procedure. As one can see from the left panel, as a result of the larger radial flow of the thermal diquarks, going from more peripheral to more central collisions the peak in the charmed baryon/meson ratio moves to higher values of $p_T$. On the other hand, integrating the respective momentum distributions, one would find that the fragmentation fractions of charm quarks into the different hadrons is independent both of the collision centrality and of the transport coefficients affecting their propagation in the deconfined fireball: hence, in the left panel of Fig.~\ref{Fig:hc-vs-pt} one simply observes a reshuffling of the charmed hadron momenta. In the right panel of Fig.~\ref{Fig:hc-vs-pt} we study the effect of hadronization on the charmed-hadron elliptic flow. Up to $p_T\approx 3$ GeV one clearly observes a mass-ordering of the $v_2$, with two distinct bands for the charmed mesons and baryons, and an inverted hierarchy at higher momenta. This behavior in our model has to be attributed to the different masses of the thermal quarks and diquarks with which recombination takes place.

\begin{figure*}
\centering
        \includegraphics[clip,height=6cm]{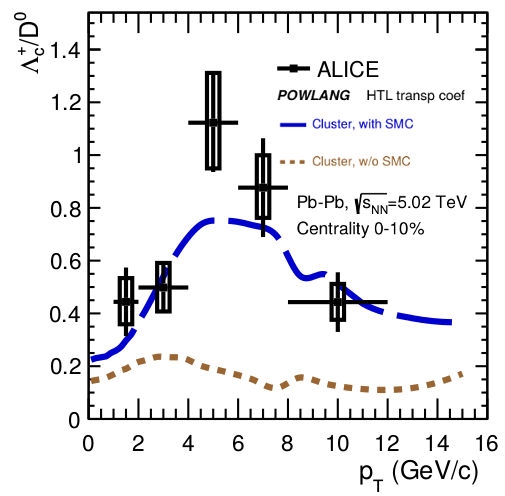}
        \includegraphics[clip,height=6cm]{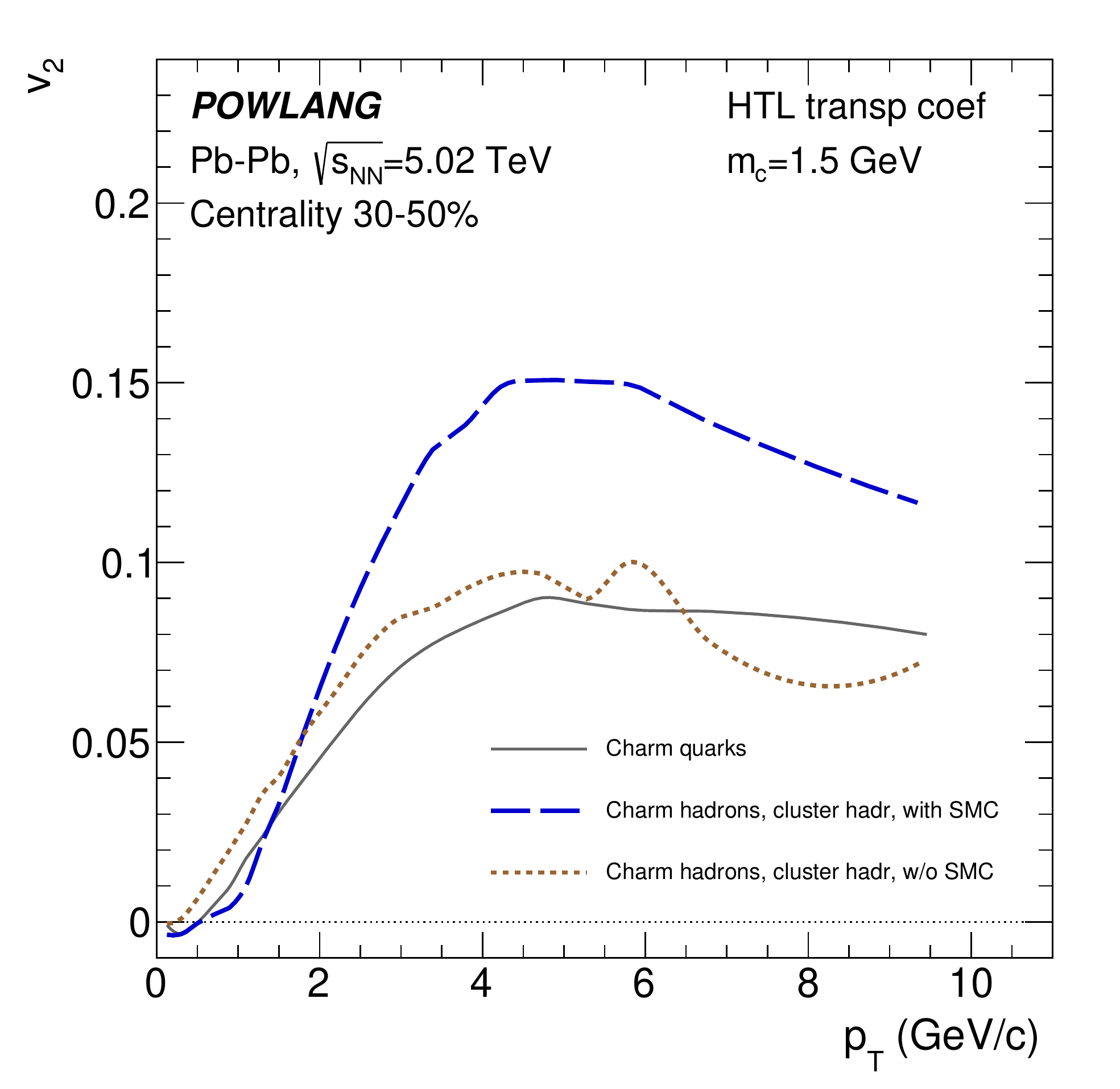}
\caption{Charmed-hadron ratios (left panel) and elliptic flow (right panel) with (dashed blue curves) and without (dotted brown curves) space-momentum correlation. The last case is obtained mixing momentum and position of heavy quarks from different events on the hadronization hypersurface.}\label{Fig:yields-noSMC}
\end{figure*}
We already discussed the relevance of SMC to reproduce the correct yields and kinematic distributions of charmed hadrons. The importance of SMC can be accessed by artificially breaking the connection between the position and the momentum of the hadronizing particles. This can be implemented by randomly redistributing the quarks undergoing recombination over the hadronization hypersurface. The effect on the final charmed hadrons is shown in Fig.~\ref{Fig:yields-noSMC}: the $\Lambda_c^+/D^0$ ratio is no longer enhanced, but one finds a value around 0.1, as in $e^+e^-$ collisions; furthermore the charmed hadron $v_2$ is lower, very similar to the one of the parent heavy quarks. Recombining partons are no longer collinear: this suppresses the collective flow of the clusters and increases their invariant mass, leading them to hadronize via string fragmentation, with no enhanced baryon production.

\begin{figure*}
\centering
\includegraphics[clip,width=0.46\textwidth]{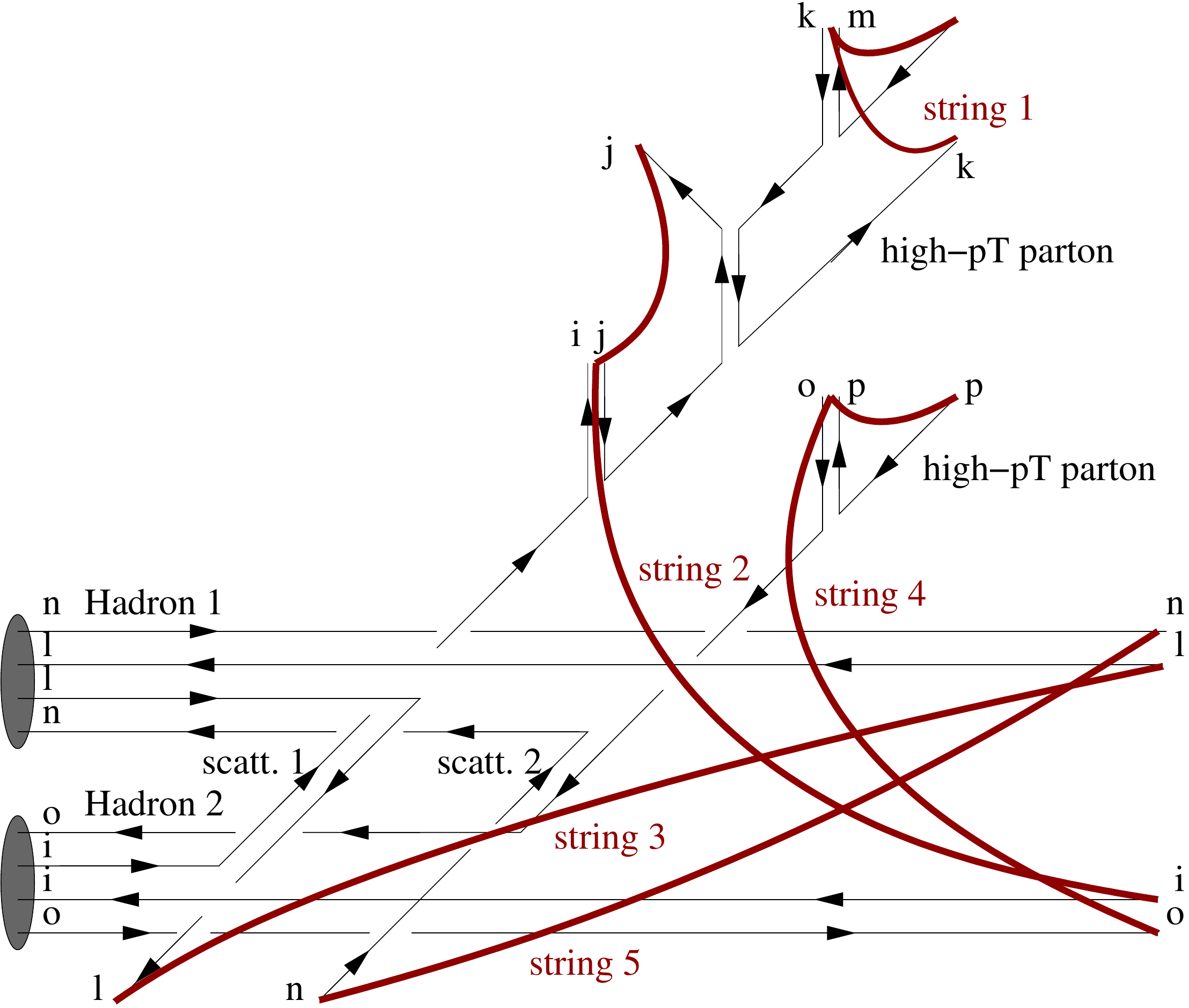}
\hspace{0.3cm}
\includegraphics[clip,width=0.46\textwidth]{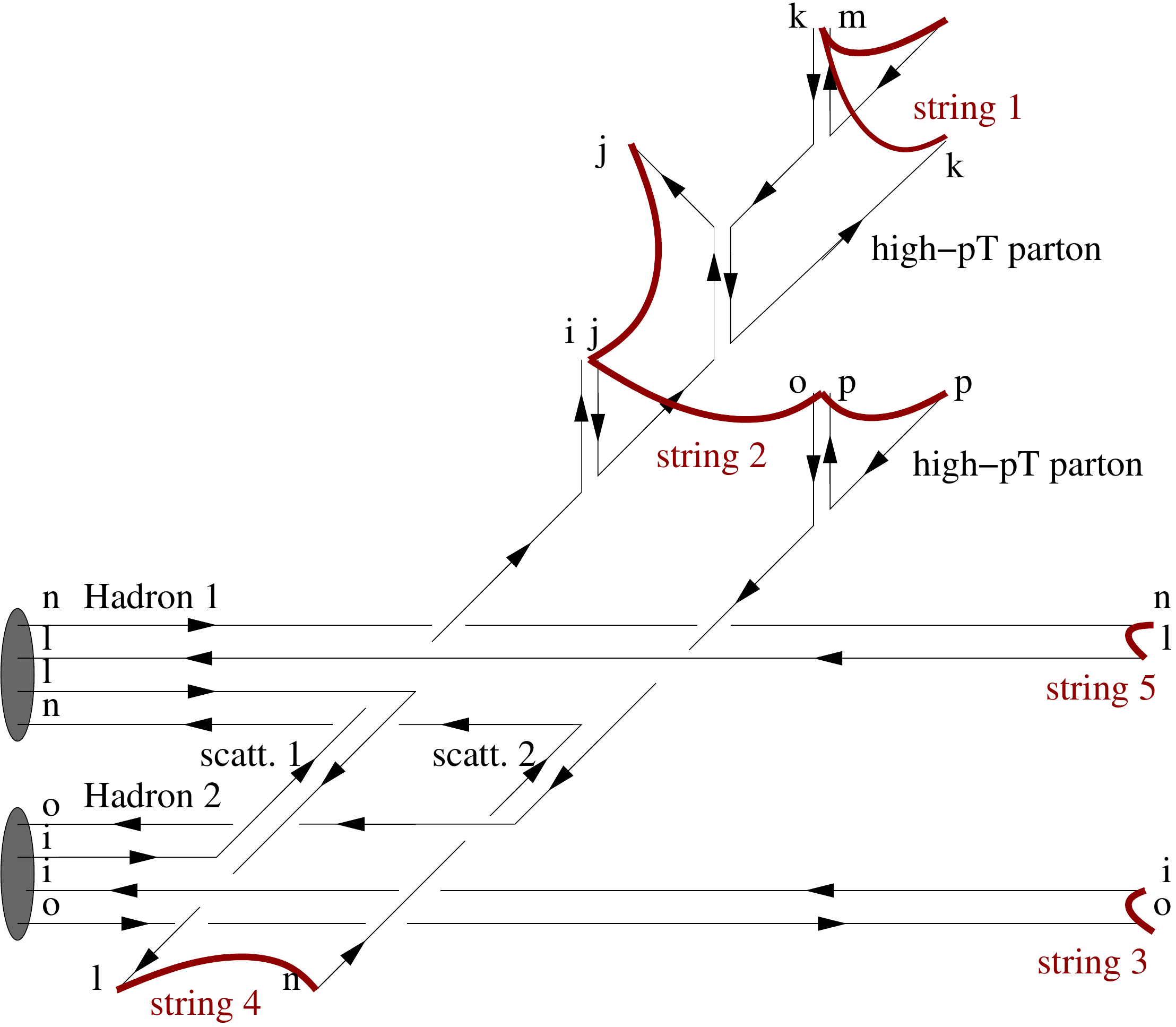}
\caption{Strings stretched between outgoing partons in a schematic hadronic collision before (left panel) and after (right panel) Color-Reconnection.}\label{Fig:CR}
\end{figure*}
Finally, we wish to discuss the common features of our hadronization mechanism with other models implemented, both recently and in the far past, in QCD event generators to reproduce experimental data. First of all, notice that the formation of low invariant-mass strings which can only decay into two particles or even collapse into a single charmed hadron was the crucial ingredient to explain, within the PYTHIA framework~\cite{Norrbin:1998bw}, asymmetries in $D^-/D^+$-meson production in $\pi^-$-proton collisions as a function of rapidity~\cite{E769:1993hmj}. Analogous studies were performed for other particle ratios and colliding systems. The only difference with our model is that charm quarks are recombined with light partons belonging to the beam remnant, while in our case they are taken from the hot medium produced in the nuclear collision.
More recently, changes in the heavy-flavor hadrochemistry in proton-proton collisions, with for example an enhanced production of charmed baryons~\cite{ALICE:2020wfu}, were interpreted as due to Color Reconnection~\cite{Christiansen:2015yqa}. The mechanism of CR is schematically illustrated in Fig.~\ref{Fig:CR}, where in the left panel we draw the strings constructed following the color-flow of the event in the large-$N_c$ approximation. However this is not necessarily the energetically most convenient configuration. Strings are extended objects and, when they are stretched in the same small region, they can overlap and interact rearranging their endpoints as shown in the right panel, if this leads to a decrease of their invariant mass. In particular, this occurs if the final strings connect more collinear partons than in the initial configuration. But this is exactly what happens in our model, which can be seen as an extreme example of CR, in which the interaction with the medium leads partons to break the color connections arising from the initial production and to recombine with nearby companions from the same fluid cell, with quite collinear momenta due to the previously discussed SMC's. Hence, the next natural step is to extend our model to describe hadronization also in proton-proton collisions, assuming that also in this case a small fireball is formed. From the latter one can extract the thermal light partons with which the charm quarks are recombined at hadronization. This is currently work in progress.  

\bibliographystyle{JHEP}
\bibliography{ICHEP-proce}

\end{document}